\documentclass[aps,pre,showpacs,reprint,amsmath,amsfonts]{revtex4-1}
\usepackage{amssymb}
\usepackage{amssymb}
\usepackage{graphicx}
\usepackage{amssymb}
\usepackage{mathtools}
\usepackage{float}
\usepackage{bbold}
\usepackage{tensor}
\usepackage{fixltx2e}
\allowdisplaybreaks

\usepackage{color}

\newcommand*\kB{\ensuremath{k_\mathrm{B}}}

\DeclareMathOperator\Imag{Im}

\newcommand{\+}{{\scriptscriptstyle{+}}}

\newcommand{\tens}[1]{\boldsymbol{#1}}
\renewcommand{\vec}[1]{\boldsymbol{#1}}

\newcommand{\caS}{{\mathcal S}}
\newcommand{\caT}{{\mathcal T}}

\newcommand{\caV}{{\mathcal V}}

\newcommand{\diff}{\text{d}}

\newcommand{\mean}[1]{{\left< #1 \right>}}
\newcommand\ul{\affiliation{Institut f\"ur Theoretische Physik, Universit\"at Leipzig,  Postfach 100 920, D-04009 Leipzig, Germany}}

\begin{document}

\title{Non-isothermal fluctuating hydrodynamics and Brownian motion}

\author{G.~Falasco}
\email{gianmaria.falasco@itp.uni-leipzig.de}
\author{K.~Kroy}
\email{klaus.kroy@uni-leipzig.de}
\ul 

\begin{abstract}
  The classical theory of Brownian dynamics follows from coarse-graining the underlying linearized fluctuating hydrodynamics of the solvent.
  We extend this procedure to globally non-isothermal conditions, requiring only a local thermal equilibration of the solvent.
  Starting from the conservation laws, we establish the stochastic
  equations of motion for the fluid momentum fluctuations in the presence of a suspended Brownian particle.
  These are then contracted to the non-isothermal generalized Langevin description of the suspended particle alone, for which the coupling to stochastic temperature fluctuations is found to be negligible under typical experimental conditions. 
\end{abstract}

 
\maketitle

\section{Introduction}

The microscopic equations of motion for strongly interacting many-body systems are, in general, intractable.
A notable exception is provided by systems exhibiting a scale separation that allows for major simplifications of these equations, making them practically (and not only formally) useful in a wide range of experimental and technological applications.
Of particular relevance is the so-called Brownian motion of a reduced number of slow degrees of freedom, for which the many fast degrees of freedom provide an effective background noise.
As Einstein realized early-on \cite{Einstein.1905}, the crucial simplification arises from the assumption that the microscopic and Brownian degrees of freedom are in thermal equilibrium, which allows for a universal characterization of the noise dynamics without explicit microscopic calculations.     
The corresponding theory of isothermal Brownian motion is by now firmly established and usually additionally exploits the fact that the mesoscopic degrees of freedom mediating between the Brownian scale and the microscopic noise degrees of freedom admit a coarse-grained, hydrodynamic description, without loss of generality.
In particular, starting with early work by Zwanzig \cite{Zwanzig.1964}, several papers have explicitly derived (generalized) Langevin equations describing Brownian motion as a contraction of the more detailed description of a fluid governed by linear fluctuating hydrodynamics \cite{Chow.1972, Hauge.1973, Bedaux.1974}.
Among the major outcomes of this inquiry there is the explanation of the long-time tails in the Brownian velocity autocorrelation function \cite{Alder.1970, Zwanzig.1970} and the robustness of the fluctuation-dissipation theorem against variations of microscopic details and even hydrodynamic specifications, such as the (in)compressibility of the solvent \cite{Chow.1973, Bedeaux.1974} or its (no-)slip boundary condition at the Brownian particle surface \cite{Bedeaux.1977}.
The necessary nanotechnological tools to conduct quantitative experimental tests of these ground-breaking theoretical developments have only become available very recently  \cite{Raizen.2010, Franosch.2011, Kheifets.2014} and vindicated the central theoretical premise ---i.e. the assumption of an underlying isothermal fluctuating solvent hydrodynamics--- with impressive precision.

Conversely, very little is known about Brownian motion in non-equilibrium solvents where the validity of a Langevin description is not \emph{a priori} ensured and standard recipes to leapfrog the microscopic dynamics using results from equilibrium statistical mechanics, such as energy equipartition, are not available.
Yet, micro- and nanoscale motion under non-equilibrium (and in particular non-isothermal) conditions is becoming increasingly relevant for innovative experimental and nanotechnological applications \cite{Sano.2009, Bechinger.2012, Qian.2013}. 

Linear fluctuating hydrodynamics, originally introduced by Landau and Lifshitz to describe density, momentum and energy fluctuations of a fluid in a global equilibrium state \cite{Landau.1987}, was later extended to non-equilibrium conditions, e.g.\
when a temperature gradient is present \cite{Ronis.1980b, Tremblay.1980, Ronis.1980, Van.1981}.
The efficacy of this non-equilibrium theory in describing fluid
fluctuations is testified by the equivalence of its predictions to
those of kinetic theory \cite{Kirk.1982}---within its range of validity, i.e.~for
dilute gases---and mode-coupling theory \cite{Procaccia.1979, Machta.1982, Kirk.1982}, and by the good agreement with light-scattering experiments; see Ref. \cite{Sengers.2006} for a review.
In view of this success, one may expect the theory to be as effective
in deriving reduced descriptions of the Brownian dynamics in
non-isothermal solvents as in the equilibrium case. 

The aim of the present work is twofold.
First, to establish the fluctuating hydrodynamic equations of motion for a non-isothermal solvent, and secondly to derive the coarse-grained description pertaining to a Brownian particle suspended therein. 
Section \ref{sec1} develops the fluctuating hydrodynamic equations suitable for addressing the Brownian motion of a sub-micron sized particle in a simple non-isothermal fluid.
More precisely, the analysis of Sec.~\ref{sec1} shows that the coupling between momentum, temperature and density gives rise to contributions that are at most proportional to $\epsilon_1 \equiv \Delta T \alpha_p$, where $\Delta T$ is the characteristic temperature variation in the system and $\alpha_p$ is the isobaric thermal expansion coefficient of the solvent.
To get a feeling for the numbers involved, consider the paradigmatic example of a hot nano-particle of radius $R\lesssim 100\,$nm in water \cite{Rings.2010}.
The temperature variations will usually be bounded by $\Delta T \lesssim 10^2$K, so that $\alpha_p\lesssim 10^{-3}\,\text{K}^{-1}$ and it is safe to assume $\epsilon_1 \ll 1$. To leading order, one can thus consider momentum and temperature fluctuations to be independent, and the fluid density to be constant.
Based on these findings, we construct the reduced description for the non-equilibrium dynamics of the immersed Brownian particle by eliminating the dynamical equations for the solvent fields, in Sec.~\ref{sec2}.
The particle position $\vec X$ turns out to evolve according to a generalized Langevin equation with long-term memory, whose zero-mean Gaussian noise satisfies a generalized fluctuation-dissipation theorem with a tensorial frequency-dependent energy spectrum $\kB \caT_{ij}(\vec X, \omega)$ that implicitly reflects the lack of homogeneity and isotropy in the fluid.
Finally, in Sec.~\ref{sec3}, we summarize our results, leaving a more thorough discussion of the consequences on the level of the coarse-grained Langevin dynamics to Ref.~\cite{Falasco.2014}.

\section{Fluctuating hydrodynamics}\label{sec1}
The starting point for the following discussion are the deterministic hydrodynamic equations describing the conservation of mass, momentum and energy in a compressible Newtonian fluid in local thermal equilibrium, which occupies the volume $\caV$ around a suspended Brownian particle of arbitrary shape.
Expressing energy in terms of the fields $T$ (local temperature) and $p$ (local pressure) by means of the local-equilibrium version of the first law we have \cite{Sengers.2006, DeGroot}
\begin{subequations}\label{hydro}
\begin{align}
& \frac{\diff \varrho}{\diff t}= - \varrho \vec \nabla \cdot\vec v, \label{mass} \\
& \varrho \frac{\diff \vec v}{\diff t}= -\vec \nabla p + \vec \nabla \cdot \vec \Gamma - \varrho g \hat{\vec z}, \label{Stokes}\\
& \varrho c_p  \frac{\diff T}{\diff t}= -\vec \nabla \cdot \vec Q - \left(\frac{\partial \varrho}{\partial T}\right)_{\!\!p} \frac{T}{\varrho} \frac{\diff p}{\diff t}, \label{heat}
\end{align}
\end{subequations}
where $\varrho$ is the mass density, $\vec v$ the velocity, $\vec \Gamma$ the deviatoric stress tensor, $\vec Q$ the heat flux, and $c_p$ the specific heat capacity at constant pressure. The gravity force $\varrho g$ is directed along the negative $z$-axis. We defined the total derivative $\frac{\diff }{\diff t}\equiv \partial_t + \vec v \cdot \vec \nabla$ to lighten the notation and note that, in Eq.~\eqref{heat}, the temperature variations caused by the viscous heating arising from the fluid motion have been discarded as second order in the fluxes and therefore negligible in comparison with $\vec Q$. The constitutive relations for the deviatoric stress tensor and the heat flux read
\begin{align}
\label{eq:constitutive}
\vec \Gamma&= \eta[ \vec \nabla  \vec v + (\vec \nabla
\vec v)^{\! \rm T}]+\eta_{\rm b}  (\vec \nabla \cdot \vec v) \vec 1, \\ 
\vec Q&= -\kappa \vec \nabla T,
\end{align}
where $\eta$, $\eta_{\rm b}$, and $\kappa$ are the dynamical shear and bulk viscosities and the heat conductivity, respectively.
We also introduce the kinematic viscosity $\nu=\eta/\varrho$ and the heat diffusivity $a_T=\kappa/(\varrho c_p)$ (the diffusion coefficients of momentum and heat), for later convenience.
We note that, at this stage, all transport coefficients can be thought of as spatially varying functions that would have to be specified, together with a material law $\varrho(p,T)$, to close the system of equations.
Having highly incompressible solvents like water in mind, we simplify the following discussion by demanding perfect incompressibility, from the outset.
Thereby, we forgo the opportunity to faithfully discuss very fast processes (faster than the time a sound wave needs to travel across a distance of about the particle size).
By moreover neglecting a possible temperature dependence of the expansion coefficient $\alpha_p$, which is again justifiable for water, the deterministic equations of motion for the solvent are closed by the simple material relation
\begin{align}\label{rho}
\varrho(\vec r, t)= \varrho_0 \Big[ 1- \alpha_p \big(T(\vec r,t)-T_0 \big)\Big],
\end{align}
where $\varrho_0\equiv\varrho(T_0)$ is the density corresponding to the reference ambient temperature $T_0$, and terms of order $O(\epsilon_1^2)$ were neglected.

The boundary condition associated to Eqs.~\eqref{Stokes}, which accounts for the momentum exchange with the suspended Brownian particle, is the no-slip condition at the particle surface $\caS$, i.e. 
\begin{equation}\label{BC}
\vec v(\vec r, t) =\mean{ \vec V(t)} + \mean{\vec \Omega(t)} \times \vec r \quad \mbox{on } \caS 
\end{equation}
where $\mean{\vec V}$ and $\mean{ \vec \Omega}$ are the deterministic translational and angular velocity of the particle, respectively \footnote{The time dependence of $\caS$ will always be neglected in the following. This amounts to move to the particle frame and drop the advection terms}. The boundary conditions for Eq.~\eqref{heat}, which describe the heat sources maintaining the inhomogeneous temperature field, as well as the particle's equations of motion, are for the moment irrelevant.

Equations~\eqref{hydro} provide the basis for describing the deterministic evolution of the coarse-grained non-equilibrium state of the fluid.
Fluctuations about this ``average state'' can be incorporated by adding stochastic terms to the stress tensor and the heat flux by substituting $\vec \Gamma \to \delta \vec \Gamma + \vec \tau$ and $\vec Q \to \delta \vec Q + \vec J$, in order to represent the random exchange of momentum and energy between the hydrodynamic and the omitted microscopic degrees of freedom \cite{Landau.1987, Sengers.2006}.
As a consequence, the hydrodynamic fields also acquire stochastic contributions according to $\vec v \to \vec v +\delta \vec v$, $p \to p +\delta p $ and $T \to T + \delta T$.
Consistency with the local-equilibrium hypothesis sets two constraints.
First, the probability densities of $\vec \tau$ and $\vec J$ must be Gaussian with mean zero and their variance obeying the local fluctuation-dissipation theorem governed by the deterministic local values of the temperature field.
Second, the stochastic equations obeyed by the fluctuating fields should be linearized, since non-linear contributions to the random fluxes are small by construction. 

We proceed as follows.
First, we identify further sub-leading terms in the deterministic hydrodynamic equations, which we simplify accordingly.
We then insert the random contributions to the stress tensor and the heat flux to obtain the corresponding stochastic equations of motion.
Our analysis closely follows the one used in the Rayleigh--B\'{e}nard problem leading to the linearized fluctuating Boussinesq equations \cite{Sengers.2006}.
However, an important difference concerns the characteristic length scale $R$ in the two problems.
Namely, in our system the relevant length scale is set by the particle size, which we assume to be at least several nanometers and less than a micron, typically on the order of $R\simeq 10^{-7}\,$m.
On this scale, advection is much less effective than diffusion in spreading momentum and heat in the fluid.
The relative magnitude of advection and diffusion terms in Eqs.~\eqref{Stokes} and \eqref{heat} is estimated as
\begin{align}
&O\left(\frac{|\vec v \cdot \vec \nabla \vec v |}{| \nu  \vec \nabla^2 \vec v|}\right) \simeq \text{Re}, && O\left(\frac{|\vec v \cdot \vec \nabla T |}{| a_T \vec \nabla^2 T|}\right) \simeq \text{Pe},
\end{align}
where $\text{Re} \equiv v R/\nu$ and $\text{Pe} \equiv \text{Re}\, \nu/a_T $ are the Reynolds and Peclet numbers associated with the particle's motion, respectively.
To remain consistent with the local equilibrium assumption, the
characteristic deterministic particle velocity (that sets the order of magnitude of the fluid velocity $v$) must remain bounded by
the thermal velocity $V_{\rm th} \sim \sqrt{\kB T/(\varrho_{\rm p}
  R^3)}$, where $R$ and $\varrho_{\rm p}$ denote the radius and mass
density of the particle, respectively. In practice, deterministic particle
velocities are usually much smaller. With $\varrho_{\rm p} \simeq \varrho_0$ this translates to $\text{Re} \simeq 10^{-3}$ and $\text{Pe} \simeq 10^{-2}$, which
implies that advection can be neglected, so that the total derivative $\diff/\diff t$ can be replaced by the partial derivative $\partial_t$ when acting on $\vec v$ and $T$, in the above equations.

Under this conditions, and after substituting Eq.~\eqref{rho} into the balance equations \eqref{mass}--\eqref{heat} the deterministic equations of motion for the solvent degrees of freedom become
\begin{widetext}
\begin{subequations}\label{hydro1}
\begin{align}
&\alpha_p \partial_t T  =  \Big[ 1- \alpha_p (T-T_0)\Big] \vec \nabla \cdot\vec v,  \\
& \varrho_0\Big[ 1- \alpha_p (T-T_0)\Big] \partial_t  \vec v=  -\vec \nabla p +\vec \nabla \cdot \vec \Gamma -  \varrho_0\Big[ 1- \alpha_p \big(T-T_0)\Big] g \hat{\vec z}, \\
& \varrho_0 c_p\Big[ 1- \alpha_p (T-T_0) \Big] \partial_t  T= -\vec \nabla \cdot \vec Q + \frac{\alpha_p T}{ 1- \alpha_p (T-T_0)} \frac{\diff p}{\diff t},
\end{align}
\end{subequations}
\end{widetext}
Notice that, in general, momentum and temperature do not evolve independently.
In order to understand the relative importance of the different terms determining such coupling we switch to dimensionless variables:
\begin{align}
& \tilde{\vec r} R \equiv r, && \tilde t \frac{R^2}{a_T} \equiv t, &&& \tilde{\vec v} \frac{a_T}{R} \equiv \vec v, \\
&\tilde T \Delta T \equiv T, && \tilde{\vec \Gamma} \frac{\varrho_0 a_T^2}{R^2}\equiv \vec \Gamma, &&& \tilde{\vec Q} \frac{ \varrho_0 c_p a_T \Delta T}{R} \equiv \vec Q.
\end{align} 
For simplicity we have taken $a_T$ and $c_p$ constant, here. Equations \eqref{hydro1} in dimensionless form are
\begin{subequations}\label{hydro2}
\begin{align}
&\epsilon_1 \partial_{\tilde t} \tilde T=  [1-\epsilon_1 (\tilde T - \tilde T_0)] \tilde{\vec \nabla} \cdot \tilde{\vec v}, \label{mass2} \\
& [1-\epsilon_1 (\tilde T - \tilde T_0) ] \partial_{\tilde t} \tilde{\vec v}= -\tilde{\vec \nabla} (\tilde p +\epsilon_2 \tilde z)+\tilde{\vec \nabla} \cdot \tilde{\vec \Gamma} + \epsilon_1 \epsilon_2 \hat{\vec  z}, \label{stokes2}\\
&[1-\epsilon_1 (\tilde T - \tilde T_0) ]   \partial_{\tilde t} \tilde T= - \tilde{\vec \nabla} \cdot \tilde{\vec Q} + \frac{\epsilon_3 \tilde T}{ 1- \epsilon_1 (\tilde T - \tilde T_0)} \frac{\diff \tilde p}{\diff \tilde t}.  \label{heat2}
\end{align}
\end{subequations}
The magnitude of the various terms can be estimated by checking the
physical values of the dimensionless parameters $\epsilon_1\equiv
\alpha_p \Delta T$,  $\epsilon_2 \equiv R^3 g/a_{T}^2$, and $\epsilon_3 \equiv \alpha_p a_T^2/(c_p
R^2)$, which control the relative magnitude of temperature-induced inhomogeneities in density, buoyancy, and pressure-driven heat fluxes, respectively. Taking  $R$ and $\Delta T$ as above, we obtain for water around
standard conditions: $\epsilon_1 \simeq 10^{-2}$, $\epsilon_2
\simeq 10^{-6}$, and $\epsilon_3 \simeq 10^{-7}$. This implies that the limit $\epsilon_1,\epsilon_2,\epsilon_3
\to 0$ reproduces the leading behavior of
Eqs.~\eqref{hydro2}, while perturbative corrections
should follow by expanding the hydrodynamic fields in series of these
small parameters. To leading order, Eqs.~\eqref{hydro2} then read
\begin{subequations}\label{hydro3}
\begin{align}
&\tilde{\vec \nabla} \cdot \tilde{\vec v}=0, \label{mass3} \\
& \partial_{\tilde t} \tilde{\vec v}= -\tilde{\vec \nabla} \tilde p +\tilde{\vec \nabla} \cdot \tilde{\vec \Gamma},  \label{stokes3}\\
&  \partial_{\tilde t} \tilde T = - \tilde{\vec \nabla} \cdot \tilde{\vec Q}. \label{heat3}
\end{align}
\end{subequations}
The condition \eqref{mass3} of a divergence-free velocity field means
that the fluid density can be treated as a constant. Restoring the physical dimensions we obtain our final set of deterministic equations
\begin{subequations}\label{det}
\begin{align}
&\vec \nabla \cdot \vec v(\vec r, t)=0,\label{div}\\ 
&\varrho_0\partial_t \vec v(\vec r, t)= -\vec \nabla  \cdot \Big[ p(\vec r, t) \vec 1 - \vec \Gamma(\vec r,t) \Big], \label{mom}\\
&\partial_t T(\vec r, t) = - a_T  \vec \nabla^2 T(\vec r, t) \label{T},
\end{align}
\end{subequations}
where the temperature dependence of the viscosity $\eta[T(\vec r, t)]$ in the deviatoric stress tensor $\vec \Gamma(\vec r ,t)$ is retained.

Now we turn to fluctuations and introduce the random stress and heat
flux into Eqs.~\eqref{mom}--\eqref{T}. The resulting fluctuating fields obey the equations
\begin{widetext}
\begin{subequations}\label{flut}
\begin{align}
&\vec \nabla \cdot \delta \vec v(\vec r, t)=0,\label{divf}\\ 
&\varrho_0\partial_t \delta \vec v(\vec r, t)= -\vec \nabla  \cdot \Big[ \delta p(\vec r, t) \vec 1 - \eta[T(\vec r, t)] \left(\vec \nabla  \vec v(\vec r, t) + \vec \nabla
\delta \vec v(\vec r, t)^{\!\rm T}\right)\Big] +\vec \nabla  \cdot \vec \tau(\vec r, t) , \label{momf}\\
&\partial_t \delta T(\vec r, t) = - a_T  \vec \nabla^2 \delta T(\vec r, t) \label{Tf} - \vec \nabla  \cdot \vec J(\vec r, t)\;.
\end{align} 
\end{subequations}
Clearly the boundary condition \eqref{BC} becomes 
\begin{equation}\label{BC2}
\delta \vec v(\vec r, t) = \delta \vec V(t) + \delta \vec \Omega(t) \times \vec r \quad \mbox{on } \caS ,
\end{equation}
where $ \delta \vec V$ and $ \delta \vec \Omega$ are the stochastic components of the particle velocities. The correlations of $\vec \tau$ and $\vec J$ are prescribed by the local-equilibrium fluctuation-dissipation relations
\begin{subequations}\label{corr}
\begin{align}\label{fluidfdt}
&\left< \tau_{ij}(\vec r, t) \tau_{kl}(\vec r^\prime, t^\prime) \right>\! =\!2 \eta[T(\vec r, t)] \kB T(\vec r, t) \delta(\vec r - \vec r^\prime) \delta (t - t^\prime) \left(\delta_{ik} \delta_{jl} + \delta_{il} \delta_{jk} \right),\\
&\left< J_i(\vec r, t) J_j(\vec r^\prime, t^\prime) \right>\! =\!2 a_T \kB T(\vec r, t)^2 \delta(\vec r - \vec r^\prime) \delta (t - t^\prime)\delta_{ij},\\
&\left< \tau_{ij}(\vec r, t) J_k(\vec r', t') \right>=0\;,
\end{align}
\end{subequations}
containing only the deterministic part of the temperature field \cite{Sengers.2006}. Consistency with the no-slip boundary condition requires that $\tens \tau = 0$ on $\caS$  \cite{Hauge.1973, Bedeaux.1977}. 

Notice that in going from Eq.~\eqref{mom} to Eq.~\eqref{momf}, we have linearized the viscous stress. Indeed, direct insertion of fluctuations in Eq.~\eqref{mom} would produce  
\begin{align}
&\eta[T+\delta T]\big(\vec \nabla (\vec v+ \delta \vec v) + (\vec \nabla (\vec v+ \delta \vec v))^{\!\rm T} \big) \simeq (\eta[T]+ \frac{\partial \eta}{\partial T}[T] \delta T) \big(\vec \nabla \vec v+ (\vec \nabla \vec v)^{\!\rm T} \big) + \eta[T] \big(\vec \nabla \delta \vec v+ (\vec \nabla \delta \vec v)^{\!\rm T} \big) \;.
\end{align}
\end{widetext}
Here we have expanded the viscosity up to first oder in $\delta T$ and dropped the manifestly nonlinear
fluctuation term proportional to $\delta T \vec \nabla \delta \vec
v$. Equation~\eqref{momf} follows by neglecting the contribution due
to temperature variations in the viscosity, which is justified by the observation
\begin{align}
O\left(\frac{|(\partial \eta/\partial T) \delta T \vec \nabla \vec v|}{|\eta \vec \nabla \delta \vec v|} \right)\sim \frac{\Delta \eta}{\eta} \frac{\sqrt{\mean{\delta T^2}}}{\Delta T} \frac{V_{\rm ex}}{V_{\rm th}} \ll 1.
\end{align}
While the typical relative viscosity variation $\Delta \eta/\eta$ are of the same order as the characteristic relative temperature variation $\Delta T/T_0$, in any coarse-graining volume consisting of $N$ solvent molecules, the spontaneous local temperature fluctuations $\sqrt{\mean{\delta T^2}}$ are small compared to $T$, typically of order $O(T/\sqrt{N})$.
Moreover, the boundary conditions with the Brownian particle Eqs.~\eqref{BC},~\eqref{BC2} fix the order of magnitude of the deterministic and fluctuating velocity fields to the typical particle velocity $V_{\rm ex}$ imposed by external forces and the particle thermal velocity $V_{\rm th}$, respectively. As noted above $V_{\rm ex} \ll V_{\rm th}$ is required to conform with the underlying local equilibrium assumption. In typical applications  $V_{\rm ex} \lesssim 10^{-6} \text{ms}^{-1}$ is indeed substantially smaller than the thermal velocity $V_{\rm th} \simeq 10^{-2}\text{ms}^{-1}$. 

We thus arrive at the important conclusion that momentum and temperature are decoupled up to corrections of order $O(\epsilon_1)$, or smaller. The reason is that only the deterministic temperature $T(\vec r, t)$ appears in Eq.~\eqref{momf}, which is fully determined by Eq.~\eqref{T}.

It is interesting, now, to go back to the starting point of the present analysis, i.e.\ the assumption of vanishing solvent compressibility $\varkappa_T=0$, that permits to assume density variations to arise from temperature heterogeneities, alone.
A rough estimation of the relative density variation due to pressure variations can be obtained as follows.
According to Eqs.~\eqref{div}--\eqref{T}, the fluid is divergence-free to leading order.
The typical magnitude of (stationary) velocity and pressure variations can thus be approximated, employing the fundamental solution of the stationary Stokes equation, by $V_{\rm th}\sim F/\eta R$ and $p_{\rm th} \sim F/R^2$.
Eliminating the thermal force $F$ exerted by the fluid, we arrive at $p_{\rm th} \sim \eta V_{\rm th}/R \simeq 1\, {\rm Nm}^{-2}$.
Thus for water we get
\begin{align}
O\left(\frac{\diff \varrho}{\varrho}\bigg|_{T}\right) \simeq O\left(\varkappa_T p_{\rm th} \right) \sim 10^{-9},
\end{align} 
which demonstrates the reliability of the assumption $\varkappa_T=0$ on time scales much longer than $R \sqrt{\varkappa_{T} \rho_0} \simeq 10^{-10} \rm{s}$.

On the basis of Eqs.~\eqref{det} and \eqref{flut}, in the next section we derive the generalized Langevin equation for the suspended Brownian particle and its associated noise spectrum.

\section{Derivation of the particle's generalized Langevin equation}\label{sec2}

We now focus on the non-equilibrium Brownian particle dynamics. The full state of the fluid-particle system is given is terms of the hydrodynamic fields and the particle coordinates, namely the center of mass position $\vec X(t)$ and the translational and rotational velocity $\vec V(t)= \mean{\vec V(t)} + \delta \vec V(t)$ and $\vec \Omega(t)= \mean{\vec \Omega(t)} + \delta \vec \Omega(t)$, respectively. The latter evolve by Newton's equations of motion
\begin{subequations}\label{newton}
\begin{eqnarray}
m \dot{\vec V} &=& \vec F + \delta \vec F + \vec{F}_{\rm ext} , \label{newton1}\\
\tens{I} \cdot \dot{\vec \Omega} &=& \vec T + \delta \vec T +\vec{T}_{\rm ext}, \label{newton2}
\end{eqnarray}
\end{subequations}
where $m$ is the mass of the particle and $\tens I$ its tensor of inertia. The deterministic force and torque exerted by the fluid are
\begin{subequations}\label{formom}
\begin{eqnarray}
\vec F(t) &=&- \int_{\caS} \tens \sigma (\vec r, t) \cdot \vec n(\vec r) \, \diff^2 r,\\
\vec T(t) &=& - \int_{\caS} \vec r \times \left(\tens \sigma (\vec r, t) \cdot \vec n(\vec r)\right) \diff^2 r,
\end{eqnarray}
\end{subequations}
where $\vec n(\vec r)$ is the inner normal vector field of the particle surface $\caS$, and
\begin{align}
\vec \sigma= - p\vec 1 + \eta \vec \Gamma
\end{align}
is the total stress tensor. Analogous definitions hold for the random force $\delta \vec F$ and torque $\delta \vec T$, replacing $\vec \sigma$ by $\delta \vec \sigma$. External forces  $\vec{F}_{\rm ext}(t)$ and torques $\vec{T}_{\rm ext}(t)$ may also be present. 
The system of Eqs.~\eqref{det},~\eqref{flut},~\eqref{newton} and
\eqref{formom}  entirely describes the evolution of the fluid and the Brownian particle. Our aim is to eliminate the equations for the  hydrodynamic fields and reduce Eqs. \eqref{newton}--\eqref{formom} to a generalized Langevin equation for the particle variables only. Thus we rewrite Eqs.~\eqref{newton} in the form
\begin{equation}\label{newton3}
\vec L \cdot \dot{\vec b} = \vec h + \delta \vec h+ \vec{f}_{\rm ext},
\end{equation}
where we combine the translational and rotational velocity into the $6$-vector $\vec b \equiv \left(\vec V, \vec \Omega\right)$, and we define the generalized tensor of inertia
\[
\tens L \equiv \begin{pmatrix} m  & 0\\ 0 & \tens I \end{pmatrix},
\]
and the generalized forces
\begin{equation}\label{newton4}
\vec h  \equiv \begin{pmatrix} \vec F  \\ \vec T  \end{pmatrix}, \quad \delta \vec h  \equiv \begin{pmatrix} \delta \vec F  \\ \delta \vec T  \end{pmatrix}, \quad \vec{f}_{\rm ext}  \equiv \begin{pmatrix} \vec{F}_{\rm ext}  \\ \vec{T}_{\rm ext}  \end{pmatrix}.
\end{equation}
By Eqs.~\eqref{mom},~\eqref{div},~\eqref{BC}, and \eqref{BC2}, $\vec v(\vec r, t)$ and $ p(\vec r, t)$ are linear functionals of $\vec b(t^\prime)$ with $- \infty < t^\prime < t$. Therefore, in view of Eq.~\eqref{formom}, the hydrodynamic forces necessarily contain a contribution which is a linear functional of $\vec b(t^\prime)$ with $- \infty < t^\prime < t$, i.e. we can write
\begin{equation}\label{sysforce}
\vec h(t) = - \int_{-\infty}^\infty \tens Z^+(\vec X, t - t^\prime) \cdot \mean{\vec b(t^\prime)} \diff t^\prime.
\end{equation}
Here $\tens Z^+(\vec X, t)$ is a $6 \times 6$ time-dependent causal friction tensor, which depends on the particle position owing to the non-constant fluid viscosity. We omit this dependence in the following. The very same reasoning applies to $\delta \vec h$, but in addition, since \eqref{momf} is a non-homogeneous equation due to the presence of the random stress $\vec \tau$, a term $\vec \xi(t)$ has to be included in order to account for contributions independent of the particle velocity. Hence $\delta \vec h$ consists of a friction term and a Langevin noise:
\begin{equation}\label{sysforce2}
\delta \vec h(t) = - \int_{-\infty}^\infty \tens Z^+(\vec X, t - t^\prime) \cdot \delta \vec b(t^\prime) \diff t^\prime + \vec \xi(t).
\end{equation}

In the subsequent derivation we shall derive the statistics of the Langevin noise $\vec \xi(t)$ and relate it to the dissipative term $\vec h(t)$. The linearity of the problem suggests to operate in the frequency space. Given a generic function of time $g(t)$, we denote its Fourier transform by $g(\omega)=\int_{-\infty}^{\infty} g(t) e^{-i \omega t} \diff t$. 
The complex conjugate of $g(\omega)$ will be denoted by $g^*(\omega)$.

In Fourier space Newton's equation \eqref{newton3} reads
\begin{equation}\label{newton3_bis}
-i \omega \tens L \cdot \vec b(\omega) = - \tens Z^{+}(\omega) \cdot  \vec b(\omega) + \vec \xi(\omega)+ {\vec f}_{ext}(\omega),
\end{equation}
where we used the Fourier transformed Eqs.~\eqref{sysforce}, \eqref{sysforce2}
\begin{subequations}\label{hforce}
\begin{eqnarray}
\vec{h}(\omega)&=&- \tens Z^{+}(\omega) \cdot \langle \vec b(\omega)\rangle, \label{hforce1}\\
\delta \vec h(\omega) &=&- \tens Z^{+}(\omega) \cdot \delta \vec b(\omega) + \vec \xi(\omega).\label{hforce2}\,
\end{eqnarray}
\end{subequations}
Note that the deterministic part of the velocity vector $\mean{\vec b(\omega)}$ is set by the external force and thus can be chosen arbitrarily.

We are now in the position to evaluate the statistics of the Langevin noise $\vec \xi(\omega)$. We proceed in three steps. First, we derive an expression for the real part of the friction tensor defined by \begin{equation}\label{symmetry}
Z_{ij}(\omega)\equiv Z_{ij}^\+(\omega)+{Z_{ij}^{\+}}^*(\omega).
\end{equation}
To evaluate the components of the friction tensor we make use of the property $Z^{+}_{ij}=Z^{+}_{ji}$, hinging only on the symmetry of the stress tensor $\sigma_{ij}=\sigma_{ji}$  \cite{Hauge.1973}. We exploit the freedom of choosing the boundary condition \eqref{BC} to select velocity vectors whose $\alpha$-th entry is the only non-zero one, and denote them by $\tensor[^{\scriptscriptstyle \alpha}]{b}{_i}(\omega)$ ---the superscript $\alpha$ will also be appended to the corresponding hydrodynamic fields. 
Second, we show that $\vec \xi(\omega)$ is a Gaussian variable with zero mean. Finally we link the noise correlation tensor $\langle \xi_i(\omega) \xi_j^*(\omega) \rangle$ to the friction tensor \eqref{symmetry}.

We wish to find an expression in terms of the solution to Eqs.~\eqref{mom} ---without formally solving the much more involved problem represented by the stochastic equations \eqref{momf}--- for the quantity
\begin{equation}\label{SymmetryZ}
Z_{ij}(\omega)\langle ^{\scriptscriptstyle \alpha}b_i(\omega)\rangle \langle ^{\scriptscriptstyle \beta}b_j^*(\omega)\rangle= Z_{\alpha \beta}(\omega)\langle b_\alpha(\omega)\rangle \langle b_\beta^*(\omega)\rangle,
\end{equation}
where the equality holds by virtue of the choice of $\vec b(\omega)$. In Eq.~\eqref{SymmetryZ} and in the following we apply the Einstein summation convention to latin indices only. Also, we suppress the function arguments where there is no risk of confusion. Equation \eqref{SymmetryZ} reads
\begin{widetext}
\begin{eqnarray}\nonumber
&& Z_{ij} \mean{ ^{\scriptscriptstyle \alpha}b_i} \mean{^{\scriptscriptstyle \beta}b^*_j}\stackrel{\eqref{symmetry}}{=} (Z_{ij}^\+ +{Z_{ij}^{\+}}^*)\mean{ ^{\scriptscriptstyle \alpha}b_i} \mean{^{\scriptscriptstyle \beta}b_j^*} \stackrel{\eqref{hforce1}}{=}-\left(^{\scriptscriptstyle \alpha}h_i\mean{^{\scriptscriptstyle \beta}b_i^*}+ \mean{ ^{\scriptscriptstyle \alpha}b_i} \phantom{\,}^{\scriptscriptstyle \beta}h_i^*\right)\label{Zsym} \\
&\stackrel{\eqref{formom},\eqref{newton4}}{=}& \langle ^{\scriptscriptstyle \beta}V_i^* \rangle  \int_{\caS} \,^{\scriptscriptstyle \alpha}\sigma_{ij} n_j \, \diff^2 r + \langle ^{\scriptscriptstyle \beta}\Omega_i^* \rangle \int_{\caS} (\vec r \times \left(^{\scriptscriptstyle \alpha}\tens \sigma \cdot \vec n\right))_i \, \diff^2 r + \langle ^{\scriptscriptstyle \alpha}V_i \rangle  \int_{\caS} \,^{\scriptscriptstyle \beta}\sigma^*_{ij} n_j\, \diff^2 r + \langle ^{\scriptscriptstyle \alpha}\Omega_i \rangle \int_{\caS} (\vec r \times \left(^{\scriptscriptstyle \beta}\tens{\sigma}^* \cdot \vec n\right))_i\,  \diff^2 r \nonumber \\
&=& \int_{\caS} \left(\langle ^{\scriptscriptstyle \beta}\vec V^* \rangle+ \langle ^{\scriptscriptstyle \beta}\vec \Omega^*\rangle \times \vec r\right)_i \! \,^{\scriptscriptstyle \alpha}\!\sigma_{ij} n_j \, \diff^2 r +\int_{\caS} \left(\langle ^{\scriptscriptstyle \alpha}\vec V \rangle+ \langle ^{\scriptscriptstyle \alpha}\vec \Omega\rangle \times \vec r\right)_i \! \,^{\scriptscriptstyle \beta}\!\sigma_{ij}^* n_j \, \diff^2 r \nonumber \\
&\stackrel{\eqref{BC}}{=}&\int_{\caS} {}^{\scriptscriptstyle \beta}v_i^* {}^{\scriptscriptstyle \alpha}\!\sigma_{ij} n_j \diff^2 r+\int_{\caS} {}^{\scriptscriptstyle \alpha}v_i {}^{\scriptscriptstyle \beta}\!\sigma_{ij}^* n_j\, \diff^2 r \nonumber \\
&=&\int_{\caV} \partial_j\left(^{\scriptscriptstyle \beta}v_i^* {}^{\scriptscriptstyle \alpha}\!\sigma_{ij}\right) \, \diff^3 r \label{div_th} +\int_{\caV} \partial_j\left(^{\scriptscriptstyle \alpha}v_i  {}^{\scriptscriptstyle \beta}\!\sigma_{ij}^*\right) \, \diff^3 r \label{divergence}   \\
&\stackrel{\eqref{mom}}{=}& \int_{\caV} \left( ^{\scriptscriptstyle \alpha}\!\sigma_{ij} \partial_j {}^{\scriptscriptstyle \beta}v_i^*+ {}^{\scriptscriptstyle \beta}\!\sigma_{ij}^* \partial_j {}^{\scriptscriptstyle \alpha}v_i\right) + i \varrho\omega \int_{\caV}  \left( ^{\scriptscriptstyle \alpha}v^*_i {}^{\scriptscriptstyle \beta}v_i- {}^{\scriptscriptstyle \beta}v^*_i {}^{\scriptscriptstyle \alpha}v_i\right) \, \diff^3 r \nonumber\\
&\stackrel{\eqref{div}}{=}& \int_{\caV} \left( {}^{\scriptscriptstyle \alpha}\Gamma_{ij} \partial_j {}^{\scriptscriptstyle \beta}v_i^*+ {}^{\scriptscriptstyle \beta}\Gamma_{ij}^* \partial_j {}^{\scriptscriptstyle \alpha}v_i\right) \, \diff^3 r + i \varrho\omega \int_{\caV}  \left( ^{\scriptscriptstyle \alpha}v^*_i {}^{\scriptscriptstyle \beta}v_i- {}^{\scriptscriptstyle \beta}v^*_i {}^{\scriptscriptstyle \alpha}v_i\right) \, \diff^3 r\nonumber\\
&=& 2 \int_{\caV} \phi_{\alpha \beta}\,\diff^3 r - 2 \varrho\omega\Imag \int_{\caV}  \phantom{\,}^{\scriptscriptstyle \alpha}v^*_i {}^{\scriptscriptstyle \beta}v_i \, \diff^3 r \label{friction},
\end{eqnarray}
where in \eqref{divergence} we used the divergence theorem and in \eqref{friction} we defined the generalized dissipation tensor 
\begin{equation}\label{phi}
\phi_{\alpha \beta}(\vec X, \vec r, \omega)\equiv  \eta(\vec r) \left(\partial_i {}^{\scriptscriptstyle \alpha}v_j(\vec r, \omega) \partial_i {}^{\scriptscriptstyle \beta}v_j^*(\vec r, \omega) + \partial_i {}^{\scriptscriptstyle \beta}v_j(\vec r,\omega)\partial_j {}^{\scriptscriptstyle \alpha}v_i^*(\vec r, \omega)\right)=\phi^*_{\beta\alpha}(\vec X, \vec r, \omega),
\end{equation}
where the $\vec X$-dependence of the hydrodynamic fields is not explicitly displayed.
Equation \eqref{friction} is valid whatever the magnitude of $^{\scriptscriptstyle \alpha}b_i$ and $^{\scriptscriptstyle \beta}b_j^*$, in particular when they are unit vectors. With this choice we have
\begin{equation}\label{cancel}
Z_{\alpha \beta}(\vec X, \omega)= 2 \int_{\caV} \phi_{\alpha \beta} \,\diff^3 r - 2 \varrho\omega\Imag \int_{\caV}  \phantom{\,}^{\scriptscriptstyle \alpha}v^*_i {}^{\scriptscriptstyle \beta}v_i \, \diff^3 r\,.
\end{equation}
Since $Z_{\alpha\beta}$ is real  by definition, Eqs.~\eqref{cancel} and \eqref{phi} imply that $\phi_{\alpha \beta}=\phi_{\beta \alpha}$. Besides, Eq.~\eqref{cancel} has to be invariant under exchange of $\alpha$ and $\beta$ owing to the symmetry $Z_{\alpha\beta}= Z_{\beta\alpha}$. Therefore one can eliminate the second term in Eq.~\eqref{cancel} and obtain
\begin{equation}\label{Zfinal}
Z_{\alpha \beta}(\vec X, \omega)=2 \int_{\caV}\phi_{\alpha \beta}(\vec X, \vec r, \omega) \, \diff^3 r.
\end{equation}

Then we turn to the random force $\vec \xi(\omega)$:
\begin{eqnarray}\nonumber
&&\xi_i \langle {}^{\scriptscriptstyle \alpha}b_i \rangle \stackrel{\eqref{hforce2}}{=}\left(\delta h_i+Z_{ij}^+ \delta b_j \right) \langle {}^{\scriptscriptstyle \alpha}b_i\rangle \stackrel{\eqref{hforce1}}{=} \delta h_i\langle {}^{\scriptscriptstyle \alpha}b_i \rangle- {}^{\scriptscriptstyle \alpha}h_j \delta b_j \nonumber \\
&\stackrel{\eqref{formom}}{=}& -\langle {}^{\scriptscriptstyle \alpha}V_i\rangle  \int_{\caS} \delta \sigma_{ij} n_j \, \diff^2 r - \langle {}^{\scriptscriptstyle \alpha}\Omega_i\rangle \int_{\caS} (\vec r \times \left( \delta \tens{\sigma} \cdot \vec n\right))_i  \,\diff^2 r +\delta V_i  \int_{\caS} {}^{\scriptscriptstyle \alpha}\!\sigma_{ij} n_j \, \diff^2 r +\delta\Omega_i \int_{\caS} (\vec r \times \left(\tens {}^{\scriptscriptstyle \alpha}\!\sigma \cdot \vec n\right))_i  \,\diff^2 r \nonumber\\
&=& -\int_{\caS} \big(\langle^{\scriptscriptstyle \alpha}\vec V \rangle + \langle {}^{\scriptscriptstyle \alpha} \vec \Omega \rangle \times \vec r\big)_i \delta \sigma_{ij}n_j \, \diff^2 r +\int_{\caS} \big(\delta \vec{V}+ \delta \vec{\Omega} \times \vec r\big)_i {}^{\scriptscriptstyle \alpha}\sigma_{ij} n_j \, \diff^2 r \nonumber \\
&\stackrel{\eqref{BC},\eqref{BC2} }{=}&-\int_{\caS} {}^{\scriptscriptstyle \alpha}v_i \delta \sigma_{ij} n_j\, \diff^2 r +\int_{\caS} \delta v_i {}^{\scriptscriptstyle \alpha}\sigma_{ij} n_j \, \diff^2 r \nonumber \\
&=&-\int_{\caV} \partial_j\left(  {}^{\scriptscriptstyle \alpha}v_i \delta \sigma_{ij}\right)\, \diff^3 r +\int_{\caV} \partial_j\left(  \delta v_i {}^{\scriptscriptstyle \alpha}\!\sigma_{ij}\right)\, \diff^3 r \label{div_th2} \\
&\stackrel{\eqref{mom}, \eqref{momf}}{=}&- \int_{\caV} \delta \sigma_{ij}\partial_j  {}^{\scriptscriptstyle \alpha}v_i \diff^3 r +\int_{\caV} {}^{\scriptscriptstyle \alpha}\sigma_{ij} \partial_j \delta v_i \diff^3 r- \int_{\caV} {}^{\scriptscriptstyle \alpha}v_i \partial_j  \tau_{ij} \diff^3 r  \label{trick} \\
&\stackrel{\eqref{div}, \eqref{divf}}{=}&- \int_{\caV} {}^{\scriptscriptstyle \alpha}v_i \partial_j \tau_{ij} \diff^3 r \nonumber \\
&=& \int_{\caV}  \tau_{ij}  \partial_j {}^{\scriptscriptstyle \alpha}v_i \diff^3 r \nonumber.
\end{eqnarray}
In \eqref{div_th2} we made use of the divergence theorem, and in \eqref{trick} of the property $ \delta\sigma_{ij} \partial_j  {}^{\scriptscriptstyle \alpha}v_i ={}^{\scriptscriptstyle \alpha}\sigma_{ij} \partial_j \delta v_i $, which is a direct consequence of the symmetry of $\vec \sigma$.
We thus have
\begin{equation}\label{noise}
\xi_i \langle {}^{\scriptscriptstyle \alpha}b_i \rangle = \xi_\alpha \langle b_\alpha\rangle=\int_{\caV} \tau_{ij}  \partial_j {}^{\scriptscriptstyle \alpha}v_i\,\diff^3 r,
\end{equation}
which shows that $\vec \xi$ is Gaussian with vanishing mean, being the integral of the deterministic quantity $\partial_j {}^{\scriptscriptstyle \alpha}v_i$ times the zero-mean Gaussian field $\tens \tau$. Hence, its correlation matrix suffices to specify the statistics  completely. Using \eqref{noise}, we determine the noise correlation:
\begin{eqnarray}
&&\langle \xi_i(\omega)  \xi_j^*(\omega')\rangle \langle {}^{\scriptscriptstyle \alpha}b_i(\omega)\rangle \langle {}^{\scriptscriptstyle \beta}b_j^*(\omega')\rangle = \langle \xi_\alpha(\omega)  \xi_\beta^*(\omega')\rangle \langle b_\alpha(\omega)\rangle \langle b_\beta^*(\omega')\rangle \label{xixibb}\\
&=& \int_{\caV} \diff^3 r' \int_{\caV} \diff^3 r  \, \partial_j {}^{\scriptscriptstyle \alpha}v_i(\vec r,\omega)  \langle  \tau_{ij}(\vec r, \omega) \tau_{kl}^*(\vec r', \omega') \rangle \partial_l {}^{\scriptscriptstyle \beta}v_k^*(\vec r',\omega') \nonumber\\
&=& 2 \kB \delta(\omega-\omega') \int_{\caV} \eta(\vec r) T(\vec r) \left(\partial_i {}^{\scriptscriptstyle \alpha}v_j(\vec r, \omega) \partial_i {}^{\scriptscriptstyle \beta}v_j^*(\vec r, \omega) + \partial_i {}^{\scriptscriptstyle \alpha}v_j(\vec r,\omega)\partial_j {}^{\scriptscriptstyle \beta}v_i^*(\vec r, \omega)\right) \, \diff^3 r \label{noisefdt} \\
&=& 2 \kB \delta(\omega-\omega') \int_{\caV}  \phi_{\alpha \beta}(\vec X, \vec r, \omega) T(\vec r) \, \diff^3 r \label{corr}.
\end{eqnarray}
In \eqref{noisefdt} we used the Fourier transform of \eqref{fluidfdt}. 
Setting the magnitude of $\langle b_\alpha\rangle$ and $\langle b_\beta^*\rangle$ to one, we finally obtain the noise correlation tensor in the form
\begin{equation}\label{corr}
\langle \xi_\alpha(\vec X, \omega)   \xi_\beta^*(\vec X, \omega')\rangle= \kB \caT_{\alpha \beta}(\vec X, \omega) Z_{\alpha \beta}(\vec X, \omega) \delta(\omega-\omega'),
\end{equation}
where $\caT_{\alpha\beta}$ is the frequency-dependent noise temperature defined by the spatial average of the temperature field $T(\vec r)$ performed with the dissipation tensor $\phi_{\alpha \beta}$
\begin{equation}\label{Tdef}
\caT_{\alpha\beta}(\vec X,\omega)\equiv\frac{\int_{\caV}  \phi_{\alpha\beta}(\vec X, \vec r, \omega) T(\vec r) \, \diff^3 r}{\int_{\caV}  \phi_{\alpha\beta}(\vec X, \vec r, \omega)\, \diff^3 r}.
\end{equation}

Summing up, we have arrived at the generalized Langevin equation for the particle velocity 
\begin{equation}\label{GLE}
\vec L \cdot \dot{\vec b}(t) =  - \int_{-\infty}^t \tens Z(\vec X, t - t^\prime) \cdot \vec b(t^\prime) \diff t^\prime + \vec \xi(t)+ \vec{f}_{ext}(t),
\end{equation}
\end{widetext}
where $\vec \xi(t)$ is a Gaussian noise having vanishing mean and correlations given by Eqs.~\eqref{corr},~\eqref{Tdef}. In principle one should consider Eq.~\eqref{T}, with the appropriate boundary conditions, in order to determine Eq.~\eqref{corr}. If the heat sources are independent of the particle motion, e.g. if they are placed at the outer boundaries of the fluid, the temperature field will generally depend on the instantaneous particle position. Nevertheless, in most practical cases the thermal conductivities of particle and solvent ($\kappa_p$ and $\kappa$, respectively) will be such that the feedback of the particle motion onto the temperature field (which is of the maximum relative strength $O(\kappa_p/\kappa-1)$ close to the particle surface) can be treated as a small correction to the overall temperature field in the solvent.
If the particle itself acts as the heat source, like in hot Brownian motion \cite{Rings.2010, PRE.2014}, the temperature field can be calculated once and for all in the particle frame ---advection terms arising from the change of frame can again be neglected.
Therefore, Eqs~\eqref{corr},~\eqref{Tdef} and \eqref{GLE} alone can be taken to entirely describe the particle dynamics.
The higher-order corrections should remain relatively small, if not negligible, for all practical purposes.

\section{Discussion}\label{sec3}

We have analyzed the fluctuating hydrodynamic equations for a Brownian particle suspended in solvents with moderate temperature gradients.
The main result obtained in Sec.~1 is that, on the scale relevant for the description of Brownian motion, it is sufficient to consider fluctuations of the solvent hydrodynamic fields around their local equilibrium state to linear order.
In particular, the solvent velocity and temperature fields are found to evolve independently.
This is traced back to two conditions.
First, heat and momentum diffusion, rather than advection, is the dominant transport mode in the fluid, as testified by the small Reynolds and Peclet numbers, namely $\text{Re}\simeq 10^{-3}$, $ \text{Pe} \simeq 10^{-2}$.
Secondly, commonly realized temperature gradients induce negligible relative density variations of order $\epsilon_1 \simeq 10^{-2}$.
The ensuing ineffectiveness of the momentum-temperature coupling has a remarkable consequence on the Langevin noise $\vec \xi$.
Namely, its non-equilibrium energy spectrum is only a result of the spatial inhomogenity of the random stress tensor $\vec \tau$, which is governed by the deterministic temperature field $T(\vec r, t)$.
Qualitatively, this could have been anticipated by recalling that the most important enhancement of fluctuations in non-equilibrium fluids is due to (nonlinear) convective couplings that dominate on long wavelengths, beyond the Brownian scale \cite{Schmitz.1985}.
Yet, the characteristic energy spectrum $\kB \caT_{ij}$ governing the Brownian noise shares with them the distinctive features of non-equilibrium fluctuations \cite{Schmitz.1988}, namely their long-ranged nature and their dependence on the mechanical transport properties of the fluid, as encoded in the dissipation tensor $\phi_{\alpha \beta}$. 

We conclude that an effective theory, which is linear in the particle-fluid momentum exchange and disregards stochastic fluctuations in the temperature field, is suitable to derive a contracted description for non-isothermal Brownian suspensions.
The consequences of this result are more fully explored in Ref.~\cite{Falasco.2014}, where the generalized Langevin equation \eqref{GLE} is derived by direct application of linear response theory to the momentum of a Brownian particle and to (locally equilibrated) distant coarse-grained solvent volume elements. 

Another important conclusion of our above calculations is that the Langevin noise $\vec \xi$ is characterized by Gaussian statistics.
Non-Gaussian contributions to the fluctuations (around the vanishing mean of $\vec \xi$) are peculiar features of out-of-equilibrium systems \cite{MFT}
that can appear only if non-linear fluctuations are retained in the hydrodynamic equations or if the local-equilibrium assumption for the fluid is violated.
Numerical simulations \cite{Chakraborty.2011, Barrat.2011} have shown that the velocity fluctuations of a hot Brownian particle clearly exhibit Gaussian distributed velocity and (when placed in a harmonic potential) position fluctuations, even under extreme heating conditions and under narrow confinement.
They thus corroborate our main result that a local-equilibrium linear fluctuating hydrodynamic theory with Gaussian noise provides a universal basis for deriving equations of motion for non-isothermal Brownian dynamics. 

The vanishing average of the Langevin noise implies that the important phenomenon of thermophoresis is absent in our discussion of non-isothermal Brownian dynamics.
To be precise, the stationary distribution for the particle position associated with Eq.~\eqref{GLE} will in general be non-uniform in space, owing to the position-dependent noise strength \cite{vanKampen.1988B}.
The corresponding weak kinematic thermodiffusion should however not be confused with what is commonly referred to as thermophoresis \cite{Bringuier.2007}.
The notion refers to an effect that cannot be derived within any generic hydrodynamic approach, not even if density variations induced by temperature gradients are considered \cite{Landau.1987}.
It is a result of non-equilibrium molecular interaction forces at the particle-fluid interface and therefore remains absent for any hydrodynamic boundary condition that does not directly couple the velocity and temperature gradient at the particle surface \cite{Goldhirsch.1983}.
Clearly, this does not preclude the introduction of thermophoresis into the hydrodynamic description by hand, e.g.\ by imposing a hydrodynamic slip velocity at the surface \cite{Anderson.1989}. 

Finally, it is worth recalling the limitations incurred by the incompressibility assumption, $\varkappa_T\to0$, made at the outset.
While it is certainly adequate to describe the particle motion averaged over the whole frequency spectrum, it is bound to break down at high frequencies, comparable to the inverse time needed by a sound wave to propagate over a distance comparable to the particle radius.
A corresponding effect that is well known in the equilibrium theory of Brownian motion is the failure of the incompressible theory to recover the mean-squared velocity, $\mean{V^2}= \kB T_0/m$, as predicted by energy-equipartition.
Instead, it predicts $\mean{V^2}= \kB T_0/M$ with the renormalized mass $M > m$ accounting for the added inertia of the solvent back-flow \cite{Hauge.1973}.
The discrepancy is straightforwardly resolved by observing that the limits $t \to 0$ and $\varkappa_T \to 0$ do not commute in the evaluation of the equal-time velocity autocorrelation function \cite{Bedeaux.1974, Zwanzig.1975}.
Accordingly, we expect that the high-frequency predictions of our theory will deviate from measurements performed with compressible solvents, but presumably only in a frequency range that is currently still difficult to access experimentally \cite{Franosch.2011, Kheifets.2014}.

\section{Acknowledgments}
We thank M. V. Gnann for fruitful discussions. We acknowledge financial support from the  Deutsche Forschungsgemeinschaft (DFG), the European Union and the Free State of Saxony, and the Leipzig School of Natural Science--Building with Molecules and Nano-objects (BuildMoNa).

\bibliography{derivation}

\begin{thebibliography}{40}%
\makeatletter
\providecommand \@ifxundefined [1]{%
 \@ifx{#1\undefined}
}%
\providecommand \@ifnum [1]{%
 \ifnum #1\expandafter \@firstoftwo
 \else \expandafter \@secondoftwo
 \fi
}%
\providecommand \@ifx [1]{%
 \ifx #1\expandafter \@firstoftwo
 \else \expandafter \@secondoftwo
 \fi
}%
\providecommand \natexlab [1]{#1}%
\providecommand \enquote  [1]{``#1''}%
\providecommand \bibnamefont  [1]{#1}%
\providecommand \bibfnamefont [1]{#1}%
\providecommand \citenamefont [1]{#1}%
\providecommand \href@noop [0]{\@secondoftwo}%
\providecommand \href [0]{\begingroup \@sanitize@url \@href}%
\providecommand \@href[1]{\@@startlink{#1}\@@href}%
\providecommand \@@href[1]{\endgroup#1\@@endlink}%
\providecommand \@sanitize@url [0]{\catcode `\\12\catcode `\$12\catcode
  `\&12\catcode `\#12\catcode `\^12\catcode `\_12\catcode `\%12\relax}%
\providecommand \@@startlink[1]{}%
\providecommand \@@endlink[0]{}%
\providecommand \url  [0]{\begingroup\@sanitize@url \@url }%
\providecommand \@url [1]{\endgroup\@href {#1}{\urlprefix }}%
\providecommand \urlprefix  [0]{URL }%
\providecommand \Eprint [0]{\href }%
\providecommand \doibase [0]{http://dx.doi.org/}%
\providecommand \selectlanguage [0]{\@gobble}%
\providecommand \bibinfo  [0]{\@secondoftwo}%
\providecommand \bibfield  [0]{\@secondoftwo}%
\providecommand \translation [1]{[#1]}%
\providecommand \BibitemOpen [0]{}%
\providecommand \bibitemStop [0]{}%
\providecommand \bibitemNoStop [0]{.\EOS\space}%
\providecommand \EOS [0]{\spacefactor3000\relax}%
\providecommand \BibitemShut  [1]{\csname bibitem#1\endcsname}%
\let\auto@bib@innerbib\@empty
\bibitem [{\citenamefont {Einstein}(1905)}]{Einstein.1905}%
  \BibitemOpen
  \bibfield  {author} {\bibinfo {author} {\bibfnamefont {A.}~\bibnamefont
  {Einstein}},\ }\href@noop {} {\bibfield  {journal} {\bibinfo  {journal} {Ann.
  Phys.}\ }\textbf {\bibinfo {volume} {322}},\ \bibinfo {pages} {549} (\bibinfo
  {year} {1905})}\BibitemShut {NoStop}%
\bibitem [{\citenamefont {Zwanzig}(1964)}]{Zwanzig.1964}%
  \BibitemOpen
  \bibfield  {author} {\bibinfo {author} {\bibfnamefont {R.}~\bibnamefont
  {Zwanzig}},\ }\href@noop {} {\bibfield  {journal} {\bibinfo  {journal} {J.
  Res. Natl. Bur. Std.(US) B}\ }\textbf {\bibinfo {volume} {68}},\ \bibinfo
  {pages} {143} (\bibinfo {year} {1964})}\BibitemShut {NoStop}%
\bibitem [{\citenamefont {Chow}\ and\ \citenamefont
  {Hermans}(1972)}]{Chow.1972}%
  \BibitemOpen
  \bibfield  {author} {\bibinfo {author} {\bibfnamefont {T.}~\bibnamefont
  {Chow}}\ and\ \bibinfo {author} {\bibfnamefont {J.}~\bibnamefont {Hermans}},\
  }\href@noop {} {\bibfield  {journal} {\bibinfo  {journal} {J. Chem. Phys.}\
  }\textbf {\bibinfo {volume} {56}},\ \bibinfo {pages} {3150} (\bibinfo {year}
  {1972})}\BibitemShut {NoStop}%
\bibitem [{\citenamefont {Hauge}\ and\ \citenamefont
  {Martin-L{\"o}f}(1973)}]{Hauge.1973}%
  \BibitemOpen
  \bibfield  {author} {\bibinfo {author} {\bibfnamefont {E.~H.}\ \bibnamefont
  {Hauge}}\ and\ \bibinfo {author} {\bibfnamefont {A.}~\bibnamefont
  {Martin-L{\"o}f}},\ }\href@noop {} {\bibfield  {journal} {\bibinfo  {journal}
  {J. Stat. Phys.}\ }\textbf {\bibinfo {volume} {7}},\ \bibinfo {pages} {259}
  (\bibinfo {year} {1973})}\BibitemShut {NoStop}%
\bibitem [{\citenamefont {Bedeaux}\ and\ \citenamefont
  {Mazur}(1974{\natexlab{a}})}]{Bedaux.1974}%
  \BibitemOpen
  \bibfield  {author} {\bibinfo {author} {\bibfnamefont {D.}~\bibnamefont
  {Bedeaux}}\ and\ \bibinfo {author} {\bibfnamefont {P.}~\bibnamefont
  {Mazur}},\ }\href@noop {} {\bibfield  {journal} {\bibinfo  {journal}
  {Physica}\ }\textbf {\bibinfo {volume} {76}},\ \bibinfo {pages} {247}
  (\bibinfo {year} {1974}{\natexlab{a}})}\BibitemShut {NoStop}%
\bibitem [{\citenamefont {Alder}\ and\ \citenamefont
  {Wainwright}(1970)}]{Alder.1970}%
  \BibitemOpen
  \bibfield  {author} {\bibinfo {author} {\bibfnamefont {B.~J.}\ \bibnamefont
  {Alder}}\ and\ \bibinfo {author} {\bibfnamefont {T.~E.}\ \bibnamefont
  {Wainwright}},\ }\href {\doibase 10.1103/PhysRevA.1.18} {\bibfield  {journal}
  {\bibinfo  {journal} {Phys. Rev. A}\ }\textbf {\bibinfo {volume} {1}},\
  \bibinfo {pages} {18} (\bibinfo {year} {1970})}\BibitemShut {NoStop}%
\bibitem [{\citenamefont {Zwanzig}\ and\ \citenamefont
  {Bixon}(1970)}]{Zwanzig.1970}%
  \BibitemOpen
  \bibfield  {author} {\bibinfo {author} {\bibfnamefont {R.}~\bibnamefont
  {Zwanzig}}\ and\ \bibinfo {author} {\bibfnamefont {M.}~\bibnamefont
  {Bixon}},\ }\href {\doibase 10.1103/PhysRevA.2.2005} {\bibfield  {journal}
  {\bibinfo  {journal} {Phys. Rev. A}\ }\textbf {\bibinfo {volume} {2}},\
  \bibinfo {pages} {2005} (\bibinfo {year} {1970})}\BibitemShut {NoStop}%
\bibitem [{\citenamefont {Chow}\ and\ \citenamefont
  {Hermans}(1973)}]{Chow.1973}%
  \BibitemOpen
  \bibfield  {author} {\bibinfo {author} {\bibfnamefont {T.}~\bibnamefont
  {Chow}}\ and\ \bibinfo {author} {\bibfnamefont {J.}~\bibnamefont {Hermans}},\
  }\href@noop {} {\bibfield  {journal} {\bibinfo  {journal} {Physica}\ }\textbf
  {\bibinfo {volume} {65}},\ \bibinfo {pages} {156} (\bibinfo {year}
  {1973})}\BibitemShut {NoStop}%
\bibitem [{\citenamefont {Bedeaux}\ and\ \citenamefont
  {Mazur}(1974{\natexlab{b}})}]{Bedeaux.1974}%
  \BibitemOpen
  \bibfield  {author} {\bibinfo {author} {\bibfnamefont {D.}~\bibnamefont
  {Bedeaux}}\ and\ \bibinfo {author} {\bibfnamefont {P.}~\bibnamefont
  {Mazur}},\ }\href@noop {} {\bibfield  {journal} {\bibinfo  {journal}
  {Physica}\ }\textbf {\bibinfo {volume} {78}},\ \bibinfo {pages} {505}
  (\bibinfo {year} {1974}{\natexlab{b}})}\BibitemShut {NoStop}%
\bibitem [{\citenamefont {Bedeaux}\ \emph {et~al.}(1977)\citenamefont
  {Bedeaux}, \citenamefont {Albano},\ and\ \citenamefont
  {Mazur}}]{Bedeaux.1977}%
  \BibitemOpen
  \bibfield  {author} {\bibinfo {author} {\bibfnamefont {D.}~\bibnamefont
  {Bedeaux}}, \bibinfo {author} {\bibfnamefont {A.}~\bibnamefont {Albano}}, \
  and\ \bibinfo {author} {\bibfnamefont {P.}~\bibnamefont {Mazur}},\
  }\href@noop {} {\bibfield  {journal} {\bibinfo  {journal} {Physica A}\
  }\textbf {\bibinfo {volume} {88}},\ \bibinfo {pages} {574} (\bibinfo {year}
  {1977})}\BibitemShut {NoStop}%
\bibitem [{\citenamefont {Li}\ \emph {et~al.}(2010)\citenamefont {Li},
  \citenamefont {Kheifets}, \citenamefont {Medellin},\ and\ \citenamefont
  {Raizen}}]{Raizen.2010}%
  \BibitemOpen
  \bibfield  {author} {\bibinfo {author} {\bibfnamefont {T.}~\bibnamefont
  {Li}}, \bibinfo {author} {\bibfnamefont {S.}~\bibnamefont {Kheifets}},
  \bibinfo {author} {\bibfnamefont {D.}~\bibnamefont {Medellin}}, \ and\
  \bibinfo {author} {\bibfnamefont {M.~G.}\ \bibnamefont {Raizen}},\
  }\href@noop {} {\bibfield  {journal} {\bibinfo  {journal} {Science}\ }\textbf
  {\bibinfo {volume} {328}},\ \bibinfo {pages} {1673} (\bibinfo {year}
  {2010})}\BibitemShut {NoStop}%
\bibitem [{\citenamefont {Franosch}\ \emph {et~al.}(2011)\citenamefont
  {Franosch}, \citenamefont {Grimm}, \citenamefont {Belushkin}, \citenamefont
  {Mor}, \citenamefont {Foffi}, \citenamefont {Forr{\'o}},\ and\ \citenamefont
  {Jeney}}]{Franosch.2011}%
  \BibitemOpen
  \bibfield  {author} {\bibinfo {author} {\bibfnamefont {T.}~\bibnamefont
  {Franosch}}, \bibinfo {author} {\bibfnamefont {M.}~\bibnamefont {Grimm}},
  \bibinfo {author} {\bibfnamefont {M.}~\bibnamefont {Belushkin}}, \bibinfo
  {author} {\bibfnamefont {F.~M.}\ \bibnamefont {Mor}}, \bibinfo {author}
  {\bibfnamefont {G.}~\bibnamefont {Foffi}}, \bibinfo {author} {\bibfnamefont
  {L.}~\bibnamefont {Forr{\'o}}}, \ and\ \bibinfo {author} {\bibfnamefont
  {S.}~\bibnamefont {Jeney}},\ }\href@noop {} {\bibfield  {journal} {\bibinfo
  {journal} {Nature}\ }\textbf {\bibinfo {volume} {478}},\ \bibinfo {pages}
  {85} (\bibinfo {year} {2011})}\BibitemShut {NoStop}%
\bibitem [{\citenamefont {Kheifets}\ \emph {et~al.}(2014)\citenamefont
  {Kheifets}, \citenamefont {Simha}, \citenamefont {Melin}, \citenamefont
  {Li},\ and\ \citenamefont {Raizen}}]{Kheifets.2014}%
  \BibitemOpen
  \bibfield  {author} {\bibinfo {author} {\bibfnamefont {S.}~\bibnamefont
  {Kheifets}}, \bibinfo {author} {\bibfnamefont {A.}~\bibnamefont {Simha}},
  \bibinfo {author} {\bibfnamefont {K.}~\bibnamefont {Melin}}, \bibinfo
  {author} {\bibfnamefont {T.}~\bibnamefont {Li}}, \ and\ \bibinfo {author}
  {\bibfnamefont {M.~G.}\ \bibnamefont {Raizen}},\ }\href@noop {} {\bibfield
  {journal} {\bibinfo  {journal} {Science}\ }\textbf {\bibinfo {volume}
  {343}},\ \bibinfo {pages} {1493} (\bibinfo {year} {2014})}\BibitemShut
  {NoStop}%
\bibitem [{\citenamefont {Jiang}\ \emph {et~al.}(2009)\citenamefont {Jiang},
  \citenamefont {Wada}, \citenamefont {Yoshinaga},\ and\ \citenamefont
  {Sano}}]{Sano.2009}%
  \BibitemOpen
  \bibfield  {author} {\bibinfo {author} {\bibfnamefont {H.-R.}\ \bibnamefont
  {Jiang}}, \bibinfo {author} {\bibfnamefont {H.}~\bibnamefont {Wada}},
  \bibinfo {author} {\bibfnamefont {N.}~\bibnamefont {Yoshinaga}}, \ and\
  \bibinfo {author} {\bibfnamefont {M.}~\bibnamefont {Sano}},\ }\href {\doibase
  10.1103/PhysRevLett.102.208301} {\bibfield  {journal} {\bibinfo  {journal}
  {Phys. Rev. Lett.}\ }\textbf {\bibinfo {volume} {102}},\ \bibinfo {pages}
  {208301} (\bibinfo {year} {2009})}\BibitemShut {NoStop}%
\bibitem [{\citenamefont {Blickle}\ and\ \citenamefont
  {Bechinger}(2012)}]{Bechinger.2012}%
  \BibitemOpen
  \bibfield  {author} {\bibinfo {author} {\bibfnamefont {V.}~\bibnamefont
  {Blickle}}\ and\ \bibinfo {author} {\bibfnamefont {C.}~\bibnamefont
  {Bechinger}},\ }\href@noop {} {\bibfield  {journal} {\bibinfo  {journal}
  {Nature Physics}\ }\textbf {\bibinfo {volume} {8}},\ \bibinfo {pages} {143}
  (\bibinfo {year} {2012})}\BibitemShut {NoStop}%
\bibitem [{\citenamefont {Qian}\ \emph {et~al.}(2013)\citenamefont {Qian},
  \citenamefont {Montiel}, \citenamefont {Bregulla}, \citenamefont {Cichos},\
  and\ \citenamefont {Yang}}]{Qian.2013}%
  \BibitemOpen
  \bibfield  {author} {\bibinfo {author} {\bibfnamefont {B.}~\bibnamefont
  {Qian}}, \bibinfo {author} {\bibfnamefont {D.}~\bibnamefont {Montiel}},
  \bibinfo {author} {\bibfnamefont {A.}~\bibnamefont {Bregulla}}, \bibinfo
  {author} {\bibfnamefont {F.}~\bibnamefont {Cichos}}, \ and\ \bibinfo {author}
  {\bibfnamefont {H.}~\bibnamefont {Yang}},\ }\href@noop {} {\bibfield
  {journal} {\bibinfo  {journal} {Chem. Sci.}\ }\textbf {\bibinfo {volume}
  {4}},\ \bibinfo {pages} {1420} (\bibinfo {year} {2013})}\BibitemShut
  {NoStop}%
\bibitem [{\citenamefont {Landau}\ and\ \citenamefont
  {Lifshitz}(1987)}]{Landau.1987}%
  \BibitemOpen
  \bibfield  {author} {\bibinfo {author} {\bibfnamefont {L.}~\bibnamefont
  {Landau}}\ and\ \bibinfo {author} {\bibfnamefont {E.}~\bibnamefont
  {Lifshitz}},\ }\href@noop {} {\emph {\bibinfo {title} {Fluid Mechanics}}},\
  \bibinfo {edition} {2nd}\ ed.,\ \bibinfo {series} {Course on Theoretical
  Physics}, Vol.~\bibinfo {volume} {6}\ (\bibinfo  {publisher} {Pergamon Press,
  Oxford},\ \bibinfo {year} {1987})\BibitemShut {NoStop}%
\bibitem [{\citenamefont {Ronis}\ \emph {et~al.}(1980)\citenamefont {Ronis},
  \citenamefont {Procaccia},\ and\ \citenamefont {Machta}}]{Ronis.1980b}%
  \BibitemOpen
  \bibfield  {author} {\bibinfo {author} {\bibfnamefont {D.}~\bibnamefont
  {Ronis}}, \bibinfo {author} {\bibfnamefont {I.}~\bibnamefont {Procaccia}}, \
  and\ \bibinfo {author} {\bibfnamefont {J.}~\bibnamefont {Machta}},\ }\href
  {\doibase 10.1103/PhysRevA.22.714} {\bibfield  {journal} {\bibinfo  {journal}
  {Phys. Rev. A}\ }\textbf {\bibinfo {volume} {22}},\ \bibinfo {pages} {714}
  (\bibinfo {year} {1980})}\BibitemShut {NoStop}%
\bibitem [{\citenamefont {Tremblay}\ \emph {et~al.}(1980)\citenamefont
  {Tremblay}, \citenamefont {Siggia},\ and\ \citenamefont
  {Arai}}]{Tremblay.1980}%
  \BibitemOpen
  \bibfield  {author} {\bibinfo {author} {\bibfnamefont {A.}~\bibnamefont
  {Tremblay}}, \bibinfo {author} {\bibfnamefont {E.~D.}\ \bibnamefont
  {Siggia}}, \ and\ \bibinfo {author} {\bibfnamefont {M.}~\bibnamefont
  {Arai}},\ }\href@noop {} {\bibfield  {journal} {\bibinfo  {journal} {Phys.
  Lett.}\ }\textbf {\bibinfo {volume} {76}},\ \bibinfo {pages} {57} (\bibinfo
  {year} {1980})}\BibitemShut {NoStop}%
\bibitem [{\citenamefont {Ronis}\ and\ \citenamefont
  {Putterman}(1980)}]{Ronis.1980}%
  \BibitemOpen
  \bibfield  {author} {\bibinfo {author} {\bibfnamefont {D.}~\bibnamefont
  {Ronis}}\ and\ \bibinfo {author} {\bibfnamefont {S.}~\bibnamefont
  {Putterman}},\ }\href {\doibase 10.1103/PhysRevA.22.773} {\bibfield
  {journal} {\bibinfo  {journal} {Phys. Rev. A}\ }\textbf {\bibinfo {volume}
  {22}},\ \bibinfo {pages} {773} (\bibinfo {year} {1980})}\BibitemShut
  {NoStop}%
\bibitem [{\citenamefont {Van~der Zwan}\ \emph {et~al.}(1981)\citenamefont
  {Van~der Zwan}, \citenamefont {Bedeaux},\ and\ \citenamefont
  {Mazur}}]{Van.1981}%
  \BibitemOpen
  \bibfield  {author} {\bibinfo {author} {\bibfnamefont {G.}~\bibnamefont
  {Van~der Zwan}}, \bibinfo {author} {\bibfnamefont {D.}~\bibnamefont
  {Bedeaux}}, \ and\ \bibinfo {author} {\bibfnamefont {P.}~\bibnamefont
  {Mazur}},\ }\href@noop {} {\bibfield  {journal} {\bibinfo  {journal} {Physica
  A}\ }\textbf {\bibinfo {volume} {107}},\ \bibinfo {pages} {491} (\bibinfo
  {year} {1981})}\BibitemShut {NoStop}%
\bibitem [{\citenamefont {Kirkpatrick}\ \emph {et~al.}(1982)\citenamefont
  {Kirkpatrick}, \citenamefont {Cohen},\ and\ \citenamefont
  {Dorfman}}]{Kirk.1982}%
  \BibitemOpen
  \bibfield  {author} {\bibinfo {author} {\bibfnamefont {T.~R.}\ \bibnamefont
  {Kirkpatrick}}, \bibinfo {author} {\bibfnamefont {E.~G.~D.}\ \bibnamefont
  {Cohen}}, \ and\ \bibinfo {author} {\bibfnamefont {J.~R.}\ \bibnamefont
  {Dorfman}},\ }\href {\doibase 10.1103/PhysRevA.26.950} {\bibfield  {journal}
  {\bibinfo  {journal} {Phys. Rev. A}\ }\textbf {\bibinfo {volume} {26}},\
  \bibinfo {pages} {950} (\bibinfo {year} {1982})}\BibitemShut {NoStop}%
\bibitem [{\citenamefont {Procaccia}\ \emph {et~al.}(1979)\citenamefont
  {Procaccia}, \citenamefont {Ronis}, \citenamefont {Collins}, \citenamefont
  {Ross},\ and\ \citenamefont {Oppenheim}}]{Procaccia.1979}%
  \BibitemOpen
  \bibfield  {author} {\bibinfo {author} {\bibfnamefont {I.}~\bibnamefont
  {Procaccia}}, \bibinfo {author} {\bibfnamefont {D.}~\bibnamefont {Ronis}},
  \bibinfo {author} {\bibfnamefont {M.}~\bibnamefont {Collins}}, \bibinfo
  {author} {\bibfnamefont {J.}~\bibnamefont {Ross}}, \ and\ \bibinfo {author}
  {\bibfnamefont {I.}~\bibnamefont {Oppenheim}},\ }\href {\doibase
  10.1103/PhysRevA.19.1290} {\bibfield  {journal} {\bibinfo  {journal} {Phys.
  Rev. A}\ }\textbf {\bibinfo {volume} {19}},\ \bibinfo {pages} {1290}
  (\bibinfo {year} {1979})}\BibitemShut {NoStop}%
\bibitem [{\citenamefont {Machta}\ and\ \citenamefont
  {Oppenheim}(1982)}]{Machta.1982}%
  \BibitemOpen
  \bibfield  {author} {\bibinfo {author} {\bibfnamefont {J.}~\bibnamefont
  {Machta}}\ and\ \bibinfo {author} {\bibfnamefont {I.}~\bibnamefont
  {Oppenheim}},\ }\href {\doibase
  http://dx.doi.org/10.1016/0378-4371(82)90185-6} {\bibfield  {journal}
  {\bibinfo  {journal} {Physica A}\ }\textbf {\bibinfo {volume} {112}},\
  \bibinfo {pages} {361 } (\bibinfo {year} {1982})}\BibitemShut {NoStop}%
\bibitem [{\citenamefont {De~Zarate}\ and\ \citenamefont
  {Sengers}(2006)}]{Sengers.2006}%
  \BibitemOpen
  \bibfield  {author} {\bibinfo {author} {\bibfnamefont {J.~M.~O.}\
  \bibnamefont {De~Zarate}}\ and\ \bibinfo {author} {\bibfnamefont {J.~V.}\
  \bibnamefont {Sengers}},\ }\href@noop {} {\emph {\bibinfo {title}
  {Hydrodynamic fluctuations in fluids and fluid mixtures}}},\ \bibinfo
  {edition} {1st}\ ed.\ (\bibinfo  {publisher} {Elsevier, Amsterdam},\ \bibinfo
  {year} {2006})\BibitemShut {NoStop}%
\bibitem [{\citenamefont {Rings}\ \emph {et~al.}(2010)\citenamefont {Rings},
  \citenamefont {Schachoff}, \citenamefont {Selmke}, \citenamefont {Cichos},\
  and\ \citenamefont {Kroy}}]{Rings.2010}%
  \BibitemOpen
  \bibfield  {author} {\bibinfo {author} {\bibfnamefont {D.}~\bibnamefont
  {Rings}}, \bibinfo {author} {\bibfnamefont {R.}~\bibnamefont {Schachoff}},
  \bibinfo {author} {\bibfnamefont {M.}~\bibnamefont {Selmke}}, \bibinfo
  {author} {\bibfnamefont {F.}~\bibnamefont {Cichos}}, \ and\ \bibinfo {author}
  {\bibfnamefont {K.}~\bibnamefont {Kroy}},\ }\href {\doibase
  10.1103/PhysRevLett.105.090604} {\bibfield  {journal} {\bibinfo  {journal}
  {Phys. Rev. Lett.}\ }\textbf {\bibinfo {volume} {105}},\ \bibinfo {pages}
  {090604} (\bibinfo {year} {2010})}\BibitemShut {NoStop}%
\bibitem [{\citenamefont {Falasco}\ \emph {et~al.}()\citenamefont {Falasco},
  \citenamefont {Gnann},\ and\ \citenamefont {Kroy}}]{Falasco.2014}%
  \BibitemOpen
  \bibfield  {author} {\bibinfo {author} {\bibfnamefont {G.}~\bibnamefont
  {Falasco}}, \bibinfo {author} {\bibfnamefont {M.~V.}\ \bibnamefont {Gnann}},
  \ and\ \bibinfo {author} {\bibfnamefont {K.}~\bibnamefont {Kroy}},\
  }\href@noop {} {\bibinfo  {journal} {http://arxiv.org/abs/1406.2116}\
  }\BibitemShut {NoStop}%
\bibitem [{\citenamefont {De~Groot}\ and\ \citenamefont
  {Mazur}(1984)}]{DeGroot}%
  \BibitemOpen
\bibfield  {journal} {  }\bibfield  {author} {\bibinfo {author} {\bibfnamefont
  {S.~R.}\ \bibnamefont {De~Groot}}\ and\ \bibinfo {author} {\bibfnamefont
  {P.}~\bibnamefont {Mazur}},\ }\href@noop {} {\emph {\bibinfo {title}
  {Non-equilibrium thermodynamics}}}\ (\bibinfo  {publisher} {Dover, New
  York},\ \bibinfo {year} {1984})\BibitemShut {NoStop}%
\bibitem [{Note1()}]{Note1}%
  \BibitemOpen
  \bibinfo {note} {The time dependence of ${\protect \mathcal S}$ will always
  be neglected in the following. This amounts to move to the particle frame and
  drop the advection terms}\BibitemShut {NoStop}%
\bibitem [{\citenamefont {Falasco}\ \emph {et~al.}(2014)\citenamefont
  {Falasco}, \citenamefont {Gnann}, \citenamefont {Rings},\ and\ \citenamefont
  {Kroy}}]{PRE.2014}%
  \BibitemOpen
  \bibfield  {author} {\bibinfo {author} {\bibfnamefont {G.}~\bibnamefont
  {Falasco}}, \bibinfo {author} {\bibfnamefont {M.~V.}\ \bibnamefont {Gnann}},
  \bibinfo {author} {\bibfnamefont {D.}~\bibnamefont {Rings}}, \ and\ \bibinfo
  {author} {\bibfnamefont {K.}~\bibnamefont {Kroy}},\ }\href {\doibase
  10.1103/PhysRevE.90.032131} {\bibfield  {journal} {\bibinfo  {journal} {Phys.
  Rev. E}\ }\textbf {\bibinfo {volume} {90}},\ \bibinfo {pages} {032131}
  (\bibinfo {year} {2014})}\BibitemShut {NoStop}%
\bibitem [{\citenamefont {Schmitz}\ and\ \citenamefont
  {Cohen}(1985)}]{Schmitz.1985}%
  \BibitemOpen
  \bibfield  {author} {\bibinfo {author} {\bibfnamefont {R.}~\bibnamefont
  {Schmitz}}\ and\ \bibinfo {author} {\bibfnamefont {E.}~\bibnamefont
  {Cohen}},\ }\href@noop {} {\bibfield  {journal} {\bibinfo  {journal} {J.
  Stat. Phys.}\ }\textbf {\bibinfo {volume} {39}},\ \bibinfo {pages} {285}
  (\bibinfo {year} {1985})}\BibitemShut {NoStop}%
\bibitem [{\citenamefont {Schmitz}(1988)}]{Schmitz.1988}%
  \BibitemOpen
  \bibfield  {author} {\bibinfo {author} {\bibfnamefont {R.}~\bibnamefont
  {Schmitz}},\ }\href@noop {} {\bibfield  {journal} {\bibinfo  {journal}
  {Physics Reports}\ }\textbf {\bibinfo {volume} {171}},\ \bibinfo {pages} {1}
  (\bibinfo {year} {1988})}\BibitemShut {NoStop}%
\bibitem [{\citenamefont {Bertini}\ \emph {et~al.}()\citenamefont {Bertini},
  \citenamefont {De~Sole}, \citenamefont {Gabrielli}, \citenamefont
  {Jona-Lasinio},\ and\ \citenamefont {Landim}}]{MFT}%
  \BibitemOpen
  \bibfield  {author} {\bibinfo {author} {\bibfnamefont {L.}~\bibnamefont
  {Bertini}}, \bibinfo {author} {\bibfnamefont {A.}~\bibnamefont {De~Sole}},
  \bibinfo {author} {\bibfnamefont {D.}~\bibnamefont {Gabrielli}}, \bibinfo
  {author} {\bibfnamefont {G.}~\bibnamefont {Jona-Lasinio}}, \ and\ \bibinfo
  {author} {\bibfnamefont {C.}~\bibnamefont {Landim}},\ }\href@noop {}
  {\bibinfo  {journal} {http://arxiv.org/abs/1404.6466}\ }\BibitemShut
  {NoStop}%
\bibitem [{\citenamefont {Chakraborty}\ \emph {et~al.}(2011)\citenamefont
  {Chakraborty}, \citenamefont {Gnann}, \citenamefont {Rings}, \citenamefont
  {Glaser}, \citenamefont {Otto},\ and\ \citenamefont
  {Kroy}}]{Chakraborty.2011}%
  \BibitemOpen
\bibfield  {journal} {  }\bibfield  {author} {\bibinfo {author} {\bibfnamefont
  {D.}~\bibnamefont {Chakraborty}}, \bibinfo {author} {\bibfnamefont {M.~V.}\
  \bibnamefont {Gnann}}, \bibinfo {author} {\bibfnamefont {D.}~\bibnamefont
  {Rings}}, \bibinfo {author} {\bibfnamefont {J.}~\bibnamefont {Glaser}},
  \bibinfo {author} {\bibfnamefont {F.}~\bibnamefont {Otto}}, \ and\ \bibinfo
  {author} {\bibfnamefont {K.}~\bibnamefont {Kroy}},\ }\href@noop {} {\bibfield
   {journal} {\bibinfo  {journal} {Eur. Phys. J.}\ }\textbf {\bibinfo {volume}
  {96}},\ \bibinfo {pages} {60009} (\bibinfo {year} {2011})}\BibitemShut
  {NoStop}%
\bibitem [{\citenamefont {Joly}\ \emph {et~al.}(2011)\citenamefont {Joly},
  \citenamefont {Merabia},\ and\ \citenamefont {Barrat}}]{Barrat.2011}%
  \BibitemOpen
  \bibfield  {author} {\bibinfo {author} {\bibfnamefont {L.}~\bibnamefont
  {Joly}}, \bibinfo {author} {\bibfnamefont {S.}~\bibnamefont {Merabia}}, \
  and\ \bibinfo {author} {\bibfnamefont {J.-L.}\ \bibnamefont {Barrat}},\
  }\href@noop {} {\bibfield  {journal} {\bibinfo  {journal} {EPL}\ }\textbf
  {\bibinfo {volume} {94}} (\bibinfo {year} {2011})}\BibitemShut {NoStop}%
\bibitem [{\citenamefont {Van~Kampen}(1988)}]{vanKampen.1988B}%
  \BibitemOpen
  \bibfield  {author} {\bibinfo {author} {\bibfnamefont {N.}~\bibnamefont
  {Van~Kampen}},\ }\href@noop {} {\bibfield  {journal} {\bibinfo  {journal}
  {Journal of Physics and Chemistry of Solids}\ }\textbf {\bibinfo {volume}
  {49}},\ \bibinfo {pages} {673} (\bibinfo {year} {1988})}\BibitemShut
  {NoStop}%
\bibitem [{\citenamefont {Bringuier}\ and\ \citenamefont
  {Bourdon}(2007)}]{Bringuier.2007}%
  \BibitemOpen
  \bibfield  {author} {\bibinfo {author} {\bibfnamefont {E.}~\bibnamefont
  {Bringuier}}\ and\ \bibinfo {author} {\bibfnamefont {A.}~\bibnamefont
  {Bourdon}},\ }\href@noop {} {\bibfield  {journal} {\bibinfo  {journal} {J.
  Non-Equilib. Thermodyn.}\ }\textbf {\bibinfo {volume} {32}},\ \bibinfo
  {pages} {221} (\bibinfo {year} {2007})}\BibitemShut {NoStop}%
\bibitem [{\citenamefont {Goldhirsch}\ and\ \citenamefont
  {Ronis}(1983)}]{Goldhirsch.1983}%
  \BibitemOpen
  \bibfield  {author} {\bibinfo {author} {\bibfnamefont {I.}~\bibnamefont
  {Goldhirsch}}\ and\ \bibinfo {author} {\bibfnamefont {D.}~\bibnamefont
  {Ronis}},\ }\href@noop {} {\bibfield  {journal} {\bibinfo  {journal} {Phys.
  Rev. A}\ }\textbf {\bibinfo {volume} {27}},\ \bibinfo {pages} {1616}
  (\bibinfo {year} {1983})}\BibitemShut {NoStop}%
\bibitem [{\citenamefont {Anderson}(1989)}]{Anderson.1989}%
  \BibitemOpen
  \bibfield  {author} {\bibinfo {author} {\bibfnamefont {J.~L.}\ \bibnamefont
  {Anderson}},\ }\href@noop {} {\bibfield  {journal} {\bibinfo  {journal} {Ann.
  Rev. Fluid Mech.}\ }\textbf {\bibinfo {volume} {21}},\ \bibinfo {pages} {61}
  (\bibinfo {year} {1989})}\BibitemShut {NoStop}%
\bibitem [{\citenamefont {Zwanzig}\ and\ \citenamefont
  {Bixon}(1975)}]{Zwanzig.1975}%
  \BibitemOpen
  \bibfield  {author} {\bibinfo {author} {\bibfnamefont {R.}~\bibnamefont
  {Zwanzig}}\ and\ \bibinfo {author} {\bibfnamefont {M.}~\bibnamefont
  {Bixon}},\ }\href@noop {} {\bibfield  {journal} {\bibinfo  {journal} {J.
  Fluid Mech.}\ }\textbf {\bibinfo {volume} {69}},\ \bibinfo {pages} {21}
  (\bibinfo {year} {1975})}\BibitemShut {NoStop}%
\end{thebibliography}%
\bibliographystyle{apsrev4-1}

\end{document}